\documentclass[aps,superscriptaddress,twocolumn,twoside,floatfix,prl,nofootinbib]{revtex4-1}
\usepackage{hyperref,graphicx,amsmath,amssymb,color}
\usepackage[normalem]{ulem}

\usepackage{amsfonts}
\usepackage{latexsym}
\usepackage{amsmath}
\usepackage{amssymb}
\usepackage{amsfonts}
\usepackage{amsthm}
\usepackage{mathrsfs}
\usepackage{natbib}
\usepackage{color,verbatim,graphics}
\usepackage{psfrag}
\DeclareMathAlphabet{\mathrsfs}{U}{rsfs}{m}{n}
\DeclareMathAlphabet{\mathpzc}{OT1}{pzc}{m}{it}
\DeclareMathAlphabet{\matheus}{U}{eus}{m}{n}
\DeclareMathAlphabet{\mathbbold}{U}{bbold}{m}{n}

\newcommand{\Tr}{\operatorname{tr}}

\newcommand{\ket}[1]{|#1\rangle}
\newcommand{\bra}[1]{\langle#1|}
\newcommand{\braket}[2]{\langle#1|#2\rangle}

\newcommand{\ie}{{\it{i.e.}~}}

\newcommand{\dani}[1]{{ #1}}

\newcommand{\US}{\mathrm{US}}
\renewcommand{\S}{\mathrm{S}}

\begin{document}

\title{Quantifying Einstein-Podolsky-Rosen steering}

\author{Paul Skrzypczyk}
\affiliation{ICFO-Institut de Ciencies Fotoniques, Mediterranean Technology Park, 08860 Castelldefels (Barcelona), Spain}

\author{Miguel Navascu\'{e}s}
\affiliation{H. H. Wills Physics Laboratory, University of Bristol$\text{,}$ Tyndall Avenue, Bristol, BS8 1TL, United Kingdom}

\author{Daniel Cavalcanti}
\affiliation{ICFO-Institut de Ciencies Fotoniques, Mediterranean Technology Park, 08860 Castelldefels (Barcelona), Spain}
\email{daniel.cavalcanti@icfo.es}

\begin{abstract}
Einstein-Podolsky-Rosen (EPR) steering is a form of bipartite quantum correlation that is intermediate between entanglement and Bell nonlocality. It allows for entanglement certification when the measurements performed by one of the parties are not characterised (or are untrusted) and has applications in quantum key distribution. Despite its foundational and applied importance, EPR steering lacks a quantitative assessment. Here we propose a way of quantifying this phenomenon and use it to study the steerability of several quantum states. In particular we show that every pure entangled state is maximally steerable, the projector onto the anti-symmetric subspace is maximally steerable for all dimensions, we provide a new example of one-way steering, and give strong support that states with positive-partial-transposition are not steerable. 
\end{abstract}

\maketitle

\emph{Introduction.---}Quantum systems display correlations that do not have a counterpart in classical physics. In the early days of quantum theory E. Schrodinger  noticed a consequence of these stronger-than-classical correlations and named it \emph{EPR steering} \cite{schrodinger}. EPR steering refers to the following phenomenon: two parties, Alice and Bob, share an entangled state $\ket{\psi_{AB}}$. By measuring her subsystem, Alice can remotely change (\ie steer) the state of Bob's subsystem in such a way that would be impossible if their systems were only classically correlated. The simplest example of steering is given by the maximally entangled state of two qubits $\ket{\phi^+}=(\ket{00}+\ket{11})/\sqrt{2}$. Alice can project Bob's system into the basis $\{\ket{a},\ket{a_{\bot}}\}$ by making a measurement of her subsystem in the conjugate basis $\{\ket{a}^*,\ket{a_{\bot}}^*\}$. As such, she can remotely prepare any state on Bob's subsystem, a feature that is impossible if they share only separable states.

EPR steering was recently given an operational interpretation as the distribution of entanglement by an untrusted party \cite{wiseman}: Alice wants to convince Bob, who does not trust her, that they share an entangled state. Bob, in order to be convinced, asks Alice to remotely prepare a collection of states of his subsystems. Alice performs her measurements (which are unknown to Bob) and communicates the results to him. By looking at the conditional states prepared by Alice, Bob is able to certify if they must have come from measurements on an entangled state. Interestingly, EPR steering is a form of quantum correlation that lies in between entanglement \cite{HorReview} and Bell nonlocality \cite{NLReview} since, on the one hand not every entangled state is steerable, and on the other hand some steerable states do not violate a Bell inequality \cite{wiseman}. Furthermore, like nonlocality, steering can be demonstrated in simple `tests', for example it is sufficient to consider only two measurements with two outcomes for Alice, preparing a collection of four states for Bob. As such, steering can be certified experimentally through the violation of steering inequalities, the analogue of Bell inequalities \cite{cavalcanti}. In fact several steering tests have been reported \cite{experimental steering,experimental steering2}, including a recent loophole-free experiment \cite{wittmann}. 

Apart from the fundamental interest in steering, there is also an applied motivation for studying and implementing it: steering allows for quantum key distribution (QKD) when one of the parties cannot trust their devices \cite{onesided}. This result opens a new venue for information-theoretic tasks based on EPR steering that are naturally suited to scenarios where only one party has trust of their device, which can naturally arise. One big advantage in this direction is that such scenarios are experimentally less demanding than fully device-independent protocols (where both of the parties distrust their devices) \cite{diqkd} and, at the same time, require less assumptions than standard quantum cryptographic scenarios.

Although our understanding of EPR steering has advanced greatly in the last few years, a fundamental question remains open: how to quantify it? Given that a quantum state can be used to demonstrate EPR steering, how `steerable' is it? In the present paper we introduce an operationally motivated method to quantify EPR steering of arbitrary finite dimensional bipartite quantum states. Our quantifier can be calculated by semidefinite programming, allowing one to explore a wide variety of quantum states and measurement scenarios.

We calculate our quantifier to several states of interest in quantum information: entangled pure states, Werner and isotropic states, and bound entangled states (with positive partial transposition (PPT)). Several interesting results follow from our analysis, such as: (i) every entangled pure state is maximally steerable; (ii) The maximally entangled version of Werner states (\ie the state described by the normalised projector onto the antisymmetric subspace) is maximally steerable, even though in dimensions larger than $2$ it is not known to violate any Bell inequality; (iii) we exhibit a new example of one-way EPR steering \cite{1way}; (iv) we provide further numerical evidence that bound entangled states are not steerable, hence supporting the extended Peres conjecture \cite{peres} recently investigated in \cite{Pus13}. Finally, we demonstrate the power of using random measurements -- in some cases they are more useful than maximally non-commuting observables (mutually unbiased basis) for detecting steering. 

\emph{EPR steering.---}Let us begin by describing in more detail the basic set-up of a steering scenario. We have two parties, Alice and Bob, one of whom is untrusted (Alice), and another of whom is trusted (Bob). Here, the meaning of `untrusted' is that we have no knowledge of either the state that Alice holds or the measurements she performs. All we know is that she can choose to perform one measurement from a set of $m$ choices, each of which has $n$ possible outcomes.  On the other hand, the meaning of `trusted' is that we have complete knowledge of the measurements Bob performs, meaning that he is able to do complete state tomography and give an exact quantum description of his system. 

A steering experiment can therefore be completely characterised by giving an `assemblage' $\{\sigma_{a|x}\}_{ax}$, the set of un-normalised states which Alice steers Bob into, given her choice of measurement $x$ and outcome $a$. The assemblage encodes both Alice's conditional probability distribution of her outcomes given her inputs, $P(a|x)=\Tr(\sigma_{a|x})$, as well as the conditional states prepared for Bob given Alice's input and outcome, $\hat{\sigma}_{a|x}= \sigma_{a|x} / P(a|x)$. All valid assemblages satisfy the consistency requirements 
\begin{equation}\label{e:consistency}
\begin{aligned}
	\sum_a \sigma_{a|x} &= \sum_a \sigma_{a|x'} \quad\quad \forall x \neq x' \\
	\Tr \sum_a \sigma_{a|x} &= 1
\end{aligned}
\end{equation}
which encode the fact that Alice cannot signal to Bob, and that without any knowledge about Alice, Bob still holds a valid quantum state. We denote this set of valid assemblages as $\Sigma^\S$.

In this scenario there is the set of ``uninteresting'' assemblages, which we shall denote the \emph{unsteerable} assemblages $\Sigma^{\US}$. These assemblages are those which can be created via classical strategies (\ie without using entanglement), and can be written in the following form (see Supplementary Material A for details)
\begin{equation} \label{e:unsteerable}
\begin{aligned}
&\sigma_{a|x} = \sum_\lambda D_\lambda(a|x)\sigma_\lambda &\forall a,x \\
\text{s.t.}& \quad\quad \Tr \sum_\lambda \sigma_\lambda = 1, \quad\quad \sigma_\lambda \geq 0 &\forall \lambda
\end{aligned}
\end{equation}
where $\lambda$ is a (classical) random variable held by Alice, $D_\lambda(a|x)$ are (the extremal) deterministic single party conditional probability distributions for Alice \footnote{i.e. the $D_\lambda(a|x)$ are the deterministic functions from the alphabet of $x$ to the alphabet of $a$. When there are $m$ inputs and $n$ outcomes, there are precisely $n^m$ such deterministic functions, and hence this is the size of the alphabet of $\lambda$.}, and $\sigma_\lambda$ are the states held by Bob. Such a model for how an unsteerable assemblage can be created is often refered to as a \emph{Local Hidden State} (LHS) model. Any assemblage that cannot be written in the form \eqref{e:unsteerable} constitutes a genuine resource in a steering scenario, and we shall call such assemblages \emph{steerable}. For such assemblages there is no classical explanation for how the different conditional states Bob holds could have been prepared by Alice. The \emph{steerability} of an assemblage can be demonstrated by the violation of steering inequalities \cite{cavalcanti}. 

Given an assemblage $\{\sigma_{a|x}\}_{ax}$ it is possible to test if it is within the set of unsteerable assemblages, \ie if $\{\sigma_{a|x}\}_{ax} \in \Sigma^\US$, with the following feasibility semidefinite program (SDP) \cite{Pus13}:
\begin{equation} \label{e:feasibility}
\begin{aligned}
\text{find}& \quad \left\{ \sigma_\lambda\right\}_{\lambda} \\
\text{s.t.}& \quad \sum_\lambda D_\lambda(a|x)\sigma_\lambda = \sigma_{a|x}  &\forall a,x \\
	& \quad \Tr \sum_\lambda \sigma_\lambda = 1, \quad\quad \sigma_\lambda \geq 0 &\forall \lambda
\end{aligned}
\end{equation}
In words, if one is able to find a set of positive semidefinite matrices $\{\sigma_\lambda\}_{\lambda}$ which satisfy all of the above constraints then the assemblage is unsteerable. If no such set can be found, then the assemblage is steerable. \dani{Notice that the same reasoning applies straighforwardly to bipartitions of multipartite states, and can then be used to investigate multipartite steering \cite{HeReid}.}

\emph{Quantifying EPR steering.---}As we have seen above, we know how to determine whether a given assemblage is a genuine resource as far as steering is concerned. However, what we desire is to be able to quantify how good a given assemblage is. The main result of this paper is to present an operationally motivated way to measure steerability, which we shall term the \emph{steerable weight}. We will show that this quantity is given by an SDP, and that there exist a wide range of scenarios where it can be tractably calculated.

The main idea behind the steerable weight is the following. We imagine that Alice, in preparing a given assemblage, will try to minimise the number of uses of a genuine steerable resource -- that is, she will prepare as frequently as possible an unsteerable assemblage, but also sometimes prepare a genuine steerable assemblage, such that on average she prepares the desired assemblage. That is, we decompose the assemblage as
\begin{equation}
	\sigma_{a|x} = \mu \sigma_{a|x}^\US + (1-\mu)\sigma_{a|x}^\S\quad\quad \forall a,x
\end{equation}
where $\sigma_{a|x}^\US$ is unsteerable, having a decomposition of the form \eqref{e:unsteerable}, while $\sigma_{a|x}^\S$ is a general steerable assemblage, satisfying condition \eqref{e:consistency}. We then ask for the maximum $\mu$ for which we can find such a decomposition. Denoting this maximum by $\mu^*$, our quantifier of steerability is the \emph{steerable weight} $SW = 1-\mu^*$, \ie the minimal amount of genuine steerable resource required to reproduce the given assemblage.

Given the above definition, and after some manipulations (see Supplementary Material B), one can show that $\mu^*$ is given by the solution to the following SDP:
\begin{equation} \label{e:weight SDP}
\begin{aligned}
\text{max}& \quad \Tr \sum_\lambda \sigma_\lambda \\
\text{s.t.}& \quad \sigma_{a|x} - \sum_\lambda D_\lambda(a|x)\sigma_\lambda \geq 0 \quad &\forall a,x \\
	& \quad \sigma_\lambda \geq 0 \quad &\forall \lambda
\end{aligned}
\end{equation}
A crucial point is that such a quantifier is given by an SDP, and as such there exist efficient numerical algorithms to evaluate it as long as the corresponding matrices involved are not too large, which happens to be the case for certain interesting scenarios, as we discuss further in the next section. Furthermore, the dual of program \eqref{e:weight SDP}, given by
\begin{equation} \label{e:weight SDP dual}
\begin{aligned}
\text{min}& \quad \Tr \sum_{ax} F_{a|x}\sigma_{a|x} \\
\text{s.t.}& \quad \openone - \sum_{ax} D_\lambda(a|x)F_{a|x} \leq 0 \quad &\forall \lambda \\
	& \quad F_{a|x} \geq 0 \quad &\forall a,x
\end{aligned}
\end{equation}
provides an additional operational meaning to the steerable weight -- as the minimal possible violation of any linear steering inequality (given here by the $F_{a|x}$) which takes only positive values, and for which all unsteerable assemblages achieve the minimum value of 1. Thus, given an assemblage, the above program provides an optimal linear steering inequality to test its steerability. Moreover it shows that linear steering inequalities are optimal. We elaborate more on this in Supplementary Material C.

\begin{figure}[t]
	\includegraphics[width=\columnwidth]{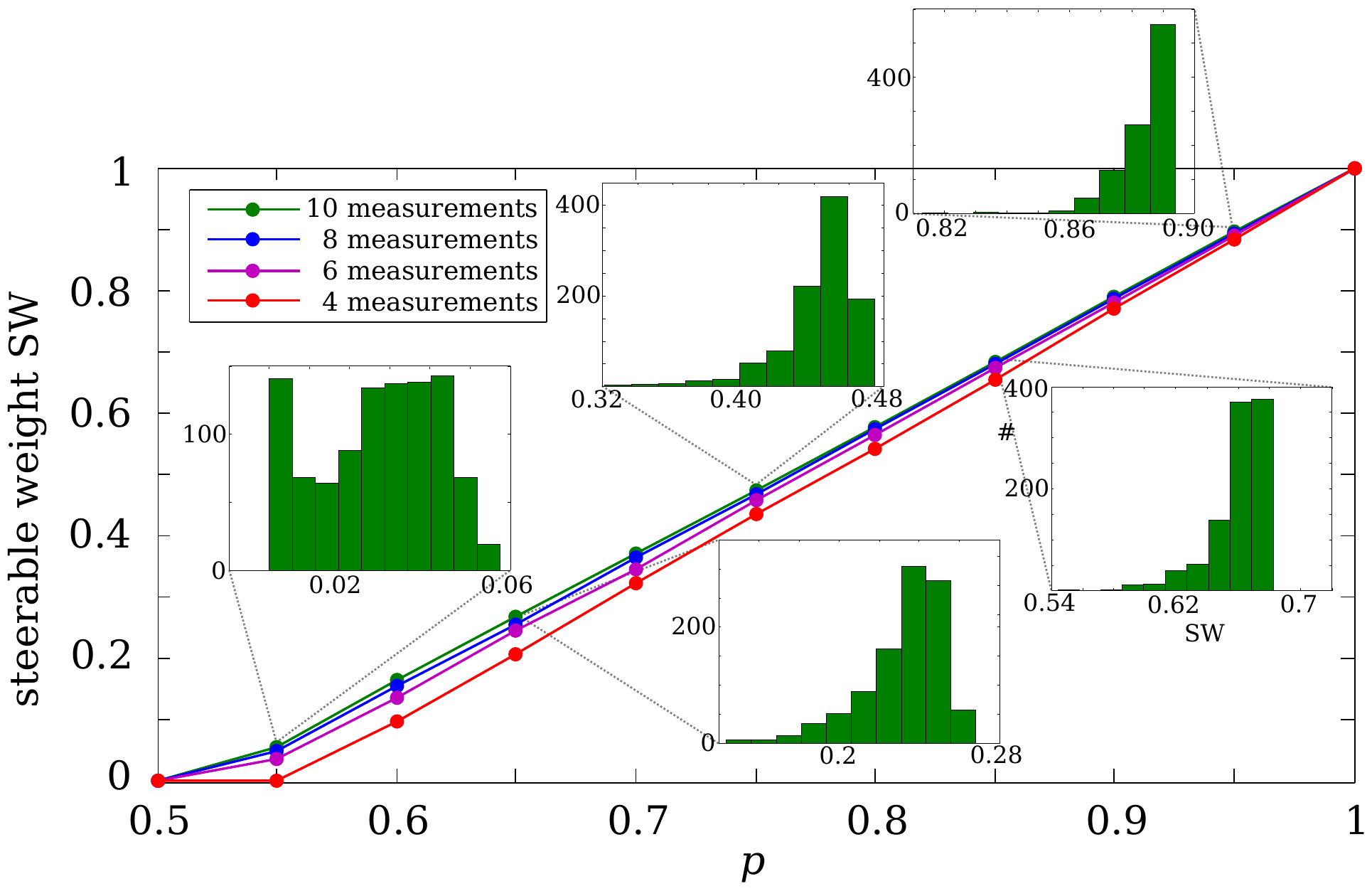} \caption{\label{f:Werner} Plot of steerable weight against parameter $p$ of 2 qubit Werner states, for varying numbers of random measurements. For each value of $p$ and a given number of measurements we plot the biggest steerable weight amongst 1000 randomly generated assemblages. Inset: For the case of 10 measurements we also show the distribution of steerable weights over the ensemble of assemblages.}
\end{figure}

\emph{Steerable weight of quantum states.---}In what follows we compute our quantifier for several examples of quantum states.

(i) \textit{Pure entangled states.} The first example we will consider are pure entangled states of arbitrary dimension $d$. In the Supplementary Material D we show that they are \emph{maximally steerable}. This can be shown by appealing to the dual characterisation of the steerable weight as given by \eqref{e:weight SDP dual}. In particular, we show that by performing two suitably chosen von Neumann measurements, Alice can create an assemblage for Bob which maximally violates an appropriately defined steering inequality. This maximal violation then implies that the assemblage is maximally steerable. 

(iia) \textit{$2 \times 2$ Werner states.} Next we consider the two-qubit Werner state $\rho =p \ket{\psi^-}\bra{\psi^-} +(1-p)\openone_2 \otimes \openone_2/4$, where $\ket{\psi^-} = \tfrac{1}{\sqrt{2}}(\ket{01}-\ket{10})$ is the singlet state and $\openone_n$ is the $n$ dimensional identity operator (in this case with $n = 2$) \cite{werner}. There are a number of different ways in which Alice can prepare assemblages for Bob, depending upon her choices of measurements. (a) Alice performs measurements of the three Pauli operators $X$, $Y$ and $Z$. We find that for the singlet ($p = 1$), the assemblage is \emph{maximally steerable} (in accordance with (i) above). As $p$ decreases, we find as expected that the steerable weight decreases monotonically, and furthermore that the assemblage becomes unsteerable (\ie has $SW = 0$) exactly when $p = 1/\sqrt{3}$, coinciding with the point where the singlet stops violating the steering inequality $\langle XX \rangle + \langle YY \rangle + \langle ZZ \rangle \leq \sqrt{3}$ \cite{cavalcanti}. (b) Alice chooses a given number $k$ of random measurements, \ie she chooses $k$ directions at random on the Bloch sphere to measure along. For $k = 4$ to $10$, we sampled over 1000 randomly generated assemblages for various values of $p$. In Fig.~\ref{f:Werner} we show, as a function of $p$, the largest steerable weight among the randomly generated assemblages, for each $k$. Firstly, for all $p$ we see that (except for the endpoints where the assemblage is either maximally steerable or completely unsteerable) as $k$ increases so does the steerable weight. Furthermore, we see that we can demonstrate steerability for Werner states with $p < 1/\sqrt{3}$ as we increase $k$, surpassing the limit for 3 measurement steering inequalities and approaching the $p=1/2$ steerable limit calculated in \cite{wiseman}. Finally, for the case of 10 measurements we also give, as insets, the distribution of steerable weight over the 1000 random assemblages for different values of $p$. We observe that as $p$ increases the distribution of steerable weights becomes increasingly peaked around the maximum value, indicating that random measurements become increasingly effective in this regime. 

(iib) \textit{d-dimensional Werner states.} We now consider the steerable weight of arbitrary dimension Werner states, defined as a convex combination of the normalized projector onto the antisymmetric subspace ($A_d$) and the normalized identity in $\mathbb{C}^d\otimes\mathbb{C}^d$ \cite{werner}:
\begin{equation}\label{werner}
\rho_W^d=\eta \frac{A_d}{N_A}+(1-\eta)\frac{\openone_{d^2}}{d^2},
\end{equation}
where $N_A=\Tr(A_d)=d(d-1)/2$.
This state is steerable if and only if $\eta>1-1/d$ \cite{wiseman}. Curiously for $d\geq3$ no Bell inequality violation is known for this state (notice that it has a LHV model for projective measurements if $\eta\leq1-1/d$ \cite{werner}). Using the steerable weight we find that for $d=3$ measuring mutually unbiase basis (MUBs) is of no use for demonstrating steering (\ie generates assemblages with $SW=0$, see Supplementary Material E for the LHS model which reproduces the assemblage), while for $d=4$ measuring MUBs demonstrates maximal steering (with $SW = 1$). However, if Alice performs $d$ random measurements onto the $d-$dimensional Werner state with $\eta=1$ she always produces a maximally steerable assmblage to Bob (\ie $SW=1$) -- see the demonstration in Supplementary Material F. This is interesting for numerous reasons. First, it contradicts the intuition that maximally non-commuting observables are the best candidates for demonstrating steering, and shows the power of randomly chosen measurements. Second, it demonstrates the existence of maximally steerable mixed states. Finally, since no Bell violation is know for Werner states with $d \geq 3$, they are good candidates for states which are maximally steerable yet Bell local. 

(iii) \textit{Erasure state and one-way steering.} We now consider the qubit erasure state $\rho^\mathrm{er}_p$,
\begin{equation}
\rho^\mathrm{er}_p = p \ket{\psi^-}\bra{\psi^-} + (1-p)\ket{2}\bra{2} \otimes \openone/2
\end{equation}
so called as it can be produced by sending Alice's half of a singlet through an erasure channel with parameter $p$, where $\ket{2}$ is the flag state. The erasure state has, for $p \leq 1/k$, a $k$--symmetric extension \cite{DPS} on Alice's side. As we show in Supplementary Material G, similarly to the case of nonlocality \cite{terhal}, it follows that if Alice performs $k$ or less measurements (including POVMs with an arbitrary number of outcomes), then there exists a LHS model, i.e. Alice cannot demonstrate steering. However, on the contrary, if we send Bob's qubit through the erasure channel, so that he holds the flag, we find that the state is steerable for all $p \neq 0$, and that this can be demonstrated with only two projective measurements for Bob (see Supplementary Material G). Thus for any arbitrary number of POVM measurements $k$ for Alice, the erasure state with $p =1/k$ is an example of a state which is unsteerable from Alice to Bob but steerable from Bob to Alice, with only the need for 2 measurements on Bob. This example complements the first demonstration of one-way steering presented in \cite{1way}, where an example was given which works for projective measurements on Alice (including the case of infinitely many measurements) and requires 13 measurements for Bob.

(iv) \textit{Bound entangled states.} Finally, we can use our quantifier to gather evidence on the Peres conjecture, that no bound entangled state can violate a Bell inequality. Since steering is a form of quantum correlation which is easier to demonstrate than nonlocality, the steerability of bound entangled states may shed light on whether one may expect them to be nonlocal also. In particular, if it is the case that bound entangled states are unsteerable then it immediately follows that they can never produce nonlocal correlations. Here we provide further numerical evidence that bound entangled states may in fact be unsteerable, and hence provide support to the Peres conjecture, which complements the recent numerical evidence given in Ref. \cite{Pus13}.

We have considered MUBs, spin, and random measurements applied to several families of bound entangled states and could not find a single instance where steering is observed.
The families of states we have explored are the (a) $3 \times 3$ unextendible product basis (UPB) states \cite{BenDivMor99}; (b) both the $3 \times 3$ Horodecki states \cite{HorHorHor98,HorHorHor01}. (c) the family of $(4,4)$ edge states\footnote{Here $(4,4)$ refers to the fact that the state is rank 4 and the partial transpose is rank 4, respectively.} of Ref.\cite{HaKyePar03}; (d) the $(5,5)$ edge state of Ref. \cite{Cla06}; (e) the family of $(5,5)$ edge states of \cite{HaKyePar03}; (f) the $(6,6)$ edge state of \cite{Ha07}; (g) the max realignment state of Ref. \cite{Cla06}; (h) The family of Bell diagonal states from \cite{Hie} for $d = 3$ and 4.

In all cases we concentrated only on cases were Alice has as many measurements as computationally feasible for the collection of statistics (in this case 6 measurements). After sampling 1000 times in each case we were unable to produce a single assemblage which was steerable. Clearly it remains to extend this approach by both collecting more data, with more measurements, and also consider more families of bound entangled states.

\emph{Comparison with entanglement and nonlocality.---}As mentioned previously, steering can be seen as an intermediate scenario which lies in between the entanglement scenario and the Bell nonlocality scenario. In the former case one trusts both parties and hence can give an exact and complete quantum description of the state $\rho_{AB}$ held by Alice and Bob. In the latter case one does not trust either party, and has only access to the measured statistics $P(ab|xy)$ related to measurement choices $x$ and $y$ of Alice and Bob, and the corresponding outcomes $a$ and $b$. 

The entanglement problem refers to deciding if a given state $\rho_{AB}$ is separable, \ie admits a decomposition of the form $\rho_{AB} = \sum_\lambda p(\lambda) \sigma_A^\lambda ~\otimes~\sigma_B^\lambda$, where $p(\lambda)$ is a probability distribution over the shared random variable $\lambda$, and $\sigma_A^\lambda$ and $\sigma_B^\lambda$ are states for Alice and Bob respectively. In the nonlocality case one is interested in deciding if a given probability distribution is local, \ie if it admits a decomposition of the form $P(ab|xy) = \sum_\lambda p(\lambda) P_\lambda(a|x) P_\lambda(b|y)$, where $p(\lambda)$ is a probability distribution over the shared random variable $\lambda$, and $P_\lambda(a|x)$ and $P_\lambda(b|y)$ are probability distributions for Alice and Bob respectively.

In all the three cases there is a way to test whether the given state, assemblage or probability distribution lies in the set of separable states, unsteerable assemblages or local distributions respectively. While we have seen that steerability can be decided using a SDP, for the case of quantum states separability can be checked by membership in a convergent hierarchy of SDPs, checking for $k$ symmetric extensions of the given state \cite{DPS} and probability distributions membership within the set of local distributions can be checked by a linear program \cite{NLReview}.  

As far as quantification is concerned, entanglement and nonlocality can also be measured by finding optimal decompositions minimising the weight on the `expensive' part. This is the so called \emph{best separable approximation} (BSA) of entangled states \cite{BSA} and the \emph{EPR2} decomposition of probability distributions \cite{EPR2}. Our results suggests that the steerable weight sometimes behaves as the BSA and sometimes as the EPR2 decomposition. For instance, every entangled two-qubit pure state is maximally entangled according to the BSA, while it is not maximally nonlocal according to the EPR2 \cite{Scarani08,PBG12}. Another (possible) difference with nonlocality is the fact that the $3 \times 3$ Werner state is steerable, while its nonlocality properties are still unknown. On the other hand, bipartite bound entangled states are conjectured to be local states (\ie with zero nonlocality according to the ERP2 decomposition). Here we find evidence that this is also true for EPR steering. 

Finally notice that although random measurements can also be used to detect nonlocality \cite{RandNL} they are not known to provide any advantage over MUBs in this case. As we have seen, random measurements can detect (even maximal) steering for cases where MUBs are useless. Furthermore they allows us to detect the steering of $2$-qubit Werner states very close to their LHS limit of $p=1/2$, then providing an interesting and scalable alternative to the previous measurement strategy based on Platonic solids \cite{experimental steering}.


\emph{Conclusion.---}In this paper we have proposed the first method to quantify the steering power of quantum states, or more precisely, of assemblages obtained by measurements on quantum states. This quantifier can be calculated using an SDP, which allowed us to estimate the steerable weight of several quantum states. We saw that the steerable weight behaves sometimes like the entanglement weight and some other times like the nonlocal weight. This confirms, in a quantitative way, that steering is an intermediate resource in-between entanglement and nonlocality. Interestingly, we have seen that mutually unbiased basis are not always the best choice of measurements to demonstrate steering.

Several questions follow from our study. Is it the case that bound entangled states are unsteerable? If this is the case then the Peres conjecture would indeed be true. Using the insight that the Peres conjecture might be even stronger than previously anticipated, could this suggest alternative ways of looking for a proof? We know that nonlocality can be superactivated, is the same also true for steering? Finally, in this study we have highlighted the power of random projective measurements. Could it be the case that going beyond projective measurements to general POVMs could provide even stronger tests of steering? 


\emph{Acknowledgements.---}This work was partially carried out at the Quantum Information 2013 conference at Benasque (Spain). We thank A. Ac\'in, N. Brunner, and L. Aolita for discussions. DC was supported by the EU project SIQS, PS by the Marie Curie COFUND action through the ICFOnest program, and M. N. by the ERC advanced grant NLST and the EPSRC DIQIP grant.

\begin{appendix}

\subsection{A: Local Hidden State Assemblages}
In this section we will show how one can arrive at the final form  
for assemblages which have a local hidden state model, given by
\begin{equation} \label{e:unsteerable}
\begin{aligned}
&\sigma_{a|x} = \sum_\lambda D_\lambda(a|x)\sigma_\lambda &\forall a,x \\
\text{s.t.}& \quad\quad \Tr \sum_\lambda \sigma_\lambda = 1, \quad\quad \sigma_\lambda \geq 0 &\forall \lambda
\end{aligned}
\end{equation}
 The most general form of an assemblage which Alice could prepare for Bob is given by
\begin{equation}\label{e:general decomp}
\sigma_{a|x} = \int p(\mu)\mathrm{d} \mu  P_\mu(a|x) \sigma_\mu
\end{equation}
where $p(\mu)\mathrm{d} \mu$ is the probability density over the hidden variable $\mu$, $P_\mu(a|x)$ is the conditional probability distribution that Alice will announce result $a$ given that her measurement was $x$ and that the hidden variable took value $\mu$, and $\sigma_\mu$ is the quantum state held by Bob when the hidden variable takes the value $\mu$. Now, for a fixed number of measurements settings and outcomes we know that any probability distribution can be written as a convex combination of deterministic strategies. That is, we can always write
\begin{equation}\label{e:deterministic decomp}
P_\mu(a|x) = \sum_\lambda p(\lambda|\mu) D_\lambda(a|x)
\end{equation} 
where $p(\lambda|\mu)$ is the weight of the deterministic strategy labelled by $\lambda$ (when the hidden variable takes value $\mu$) and the $D_\lambda(a|x)$ are the deterministic conditional probability distributions which for each choice of measurement provide a deterministic outcome. (In more detail, if the scenario at hand has $n$ possible measurements with $m$ possible outcomes, then each deterministic strategy can be thought of as an $n$-component vectors $\vec{\nu} = (\nu_1, \cdots, \nu_n)$, with all entries in the set $\{1, \cdots, m\}$, such that $P(a|x) = \delta_{a,\nu_x}$. There are $m^n$ such vectors, and thus the same number of deterministic strategies, hence $\lambda$ is simply the label of a given vector). Using \eqref{e:deterministic decomp} we can rewrite \eqref{e:general decomp} as
\begin{eqnarray}
\sigma_{a|x} &=& \int p(\mu)\mathrm{d} \mu  \sum_\lambda p(\lambda|\mu) D_\lambda(a|x) \sigma_\mu \nonumber \\
			&=& \sum_\lambda D_\lambda(a|x) \int p(\mu)\mathrm{d} \mu   p(\lambda|\mu)  \sigma_\mu \nonumber \\
			&=& \sum_\lambda D_\lambda(a|x) \tilde{\sigma}_\lambda
\end{eqnarray}
where we have defined $\tilde{\sigma}_\lambda := \int p(\mu)\mathrm{d} \mu   p(\lambda|\mu)  \sigma_\mu$. This reproduces the first line of equation \eqref{e:unsteerable} and demonstrates that without loss of generality we can restrict to local hidden state models consisting of only a finite number of hidden states, one corresponding to each deterministic strategy Alice may use to generate her output $a$ given her input $x$. 

To recover the second line of \eqref{e:unsteerable}, we can study the properties that $\tilde{\sigma}_\lambda$ inherits through its definition: (i)~Since $\sigma_\mu \geq 0$ (i.e. it is a positive semidefinite operator), and $p(\mu)\mathrm{d} \mu   p(\lambda|\mu)$ is a positive real number, then $\tilde{\sigma}_\lambda \geq 0$. (ii) Since $\Tr(\sigma_\mu) = 1$, then
\begin{eqnarray}
\Tr(\tilde{\sigma}_\lambda)  &=& \Tr\int q(\mu)\mathrm{d} \mu   p(\lambda|\mu)  \sigma_\mu \nonumber \\
						&=& \int q(\mu)\mathrm{d} \mu   p(\lambda|\mu) \nonumber \\
						&=& q(\lambda)
\end{eqnarray}
i.e. $\tilde{\sigma}_\lambda$ is a sub-normalised state whose trace gives the probability of the corresponding hidden variable $\lambda$. (iii) Since we must have $\sum_\lambda q(\lambda) = 1$, this translates to $\Tr \sum_\lambda q(\lambda) = 1$. This thus reproduces in full the definition \eqref{e:unsteerable}. 
\newline


\subsection{B: Deriving the SDP for steerable weight}
In this appendix we show how to arrive at the SDP (5) in the main text. 
First, let us write down the optimisation problem which follows directly from the definition of the steerable weight, 
\begin{equation} \label{e:weight start}
\begin{aligned}
\text{max}& \quad\mu \\
\text{s.t.}&\quad \sigma_{a|x} = \mu \sigma_{a|x}^\US + (1-\mu)\sigma_{a|x}^\S\quad \forall a,x \\
	& \quad\big\{ \sigma_{a|x}^\US\big\}_{ax} \in \Sigma^\US,\quad\big\{ \sigma_{a|x}^\S\big\}_{ax} \in \Sigma^\S
\end{aligned}
\end{equation}
This can be written more explicitly, using the definitions of the sets $\Sigma^\US$ and $\Sigma^\S$, as
\begin{equation} \label{e:weight SDP2}
\begin{aligned}
\text{max}& \quad \mu \\
\text{s.t.}& \quad \sigma_{a|x} = \mu \sigma_{a|x}^\US + (1-\mu)\sigma_{a|x}^\S &\forall a,x \\
			& \quad \sigma_{a|x}^\US =  \sum_\lambda D_\lambda(a|x)\sigma_\lambda   &\forall a,x \\
	& \quad \Tr \sum_\lambda \sigma_\lambda = 1, \quad\quad \sigma_\lambda \geq 0  &\forall \lambda \\
	& \quad \sum_a \sigma_{a|x}^\S = \sum_a \sigma_{a|x'}^\S & \forall x \neq x' \\
	& \quad \sigma_{a|x}^\S \geq 0  &\forall a,x
\end{aligned}
\end{equation}
which is still not yet in the form of an SDP. The next step is to realise that the optimisation variables $\{\sigma_{a|x}^\S\}_{ax}$ can in fact be eliminated. The last constraint, $\sigma_{a|x}^\S \geq 0$ can be re-expressed, using the definition of $\sigma_{a|x}$ and $\sigma_{a|x}^\US$ as
\begin{equation}
	\sigma_{a|x}^\S = \tfrac{1}{1-\mu}\left(\sigma_{a|x} - \sum_\lambda D_\lambda(a|x)\sigma_\lambda \right) \geq 0
\end{equation}
assuming that $\mu < 1$ (\ie that $\sigma_{a|x}^\S$ does not vanish from the problem), then the term inside the brackets must be positive semidefinite. On the other hand, the constraint $\sum_a \sigma_{a|x}^\S = \sum_a \sigma_{a|x'}^\S$ is always satisfied, as long as the input assemblage $\{\sigma_{a|x}\}_{ax}$ is consistent, satisfying conditions 
\begin{equation}\label{e:consistency}
\begin{aligned}
	\sum_a \sigma_{a|x} &= \sum_a \sigma_{a|x'} \quad\quad \forall x \neq x' \\
	\Tr \sum_a \sigma_{a|x} &= 1.
\end{aligned}
\end{equation}
In particular, we have that $\sum_a \sigma_{a|x} = \sum_a \sigma_{a|x'}$ and furthermore, since $\sum_a D_\lambda(a|x) = 1$ for all $x$ (as they are valid probability distributions), we have that
\begin{equation}
	\sum_a \sum_\lambda D_\lambda(a|x)\sigma_\lambda = \sum_\lambda \sigma_\lambda = \sigma_R
\end{equation} 
the reduced state on Bob's side, which is manifestly independent of $x$. Thus we see that all the constraints on $\sigma_{a|x}^\S$ can either be re-expressed in terms of other optimisation variables, or are automatically satisfied, and hence these variables can be eliminated from the problem.
The only variables left at therefore $\sigma_\lambda$. Defining new variables $\tilde{\sigma}_\lambda = \mu \sigma_\lambda$, we see that $\Tr \sum_\lambda \tilde{\sigma}_\lambda = \mu$, and that since $\mu \geq 0$ also $\tilde{\sigma}_\lambda \geq 0$. Combining this with the above, we finally arrive at the final form for the SDP: 
\begin{equation} \label{e:weight SDP simple}
\begin{aligned}
\text{max}& \quad \Tr \sum_\lambda \sigma_\lambda \\
\text{s.t.}& \quad \sigma_{a|x} - \sum_\lambda D_\lambda(a|x)\sigma_\lambda \geq 0 \quad \forall a,x \\
	& \quad \sigma_\lambda \geq 0 \quad \forall \lambda
\end{aligned}
\end{equation}
where for convenience we have written $\tilde{\sigma}$ as $\sigma$.


\subsection{C: Dual SDP for steerable weight: bounding $SW$ by steering inequality violations.}
Introducing dual variables $F_{a|x}$ and $G_\lambda$, dual to the first and second set of constraints respectively, one straightforwardly arrives at the following dual program to \eqref{e:weight SDP simple},
\begin{equation} \label{e:weight SDP dual full}
\begin{aligned}
\text{min}& \quad \Tr \sum_{ax} F_{a|x}\sigma_{a|x} \\
\text{s.t.}& \quad \openone - \sum_{ax} D_\lambda(a|x)F_{a|x} + G_\lambda = 0  &\forall \lambda \\
	& \quad F_{a|x} \geq 0 \quad \forall a,x \quad\quad\quad G_\lambda \geq 0  &\forall \lambda 
\end{aligned}
\end{equation}
However, the $G_\lambda$ are seen to play the role of slack variables, and therefore the dual can be further simplified to
\begin{equation}\label{e:weight SDP dual final}
\begin{aligned}
\text{min}& \quad \Tr \sum_{ax} F_{a|x}\sigma_{a|x} \\
\text{s.t.}& \quad \openone - \sum_{ax} D_\lambda(a|x)F_{a|x} \leq 0 \quad &\forall \lambda \\
	& \quad F_{a|x} \geq 0 \quad &\forall a,x
\end{aligned}
\end{equation}
The meaning of this dual formulation can easily be deduced. The operators $\{F_{a|x}\}_{ax}$ define a linear steering inequality, with the objective function $\Tr\sum_{ax}F_{a|x}\sigma_{a|x}$ being precisely the value obtained by the input assemblage $\{\sigma_{a|x}\}$. The condition that $F_{a|x} \geq 0$ enforces that the minimum value of any assemblage is non-negative. Finally, any collection of states $\{\sigma_\lambda\}_\lambda$ such that $\Tr\sum_\lambda \sigma_\lambda = 1$ and $\sigma_\lambda \geq 0$ constitutes a LHS model. Multiplying  $\openone - \sum_{ax} D_\lambda(a|x)F_{a|x}$ by $\sigma_\lambda$, taking the sum and trace, we find that 
\begin{equation}
1 \leq \Tr\sum_{ax} \sigma_{a|x}^\US F_{a|x}
\end{equation}
where we used the fact that $\sum_\lambda D_\lambda(a|x)\sigma_\lambda = \sigma_{a|x}^\US$ is an unsteerable assemblage. Therefore, this condition says that the steering inequality $\{F_{a|x}\}_{ax}$ is lower bounded by unity on all unsteerable assemblages. \dani{Thus, given an assemblage $\{\sigma_{a|x}\}$, the dual program \eqref{e:weight SDP dual final} searches for the steering inequality for which it provides the maximal violation.}

Note furthermore that the dual is seen to be strictly feasible, demonstrated by choosing the trivial inequality $F_{a|x} = \alpha \openone$ for an appropriately large $\alpha$. Thus strong duality holds and the steerable weight has a dual operational interpretation as one minus the violation of an optimal steering inequality, where inequalities are constrained to have the standardised form as required by the dual SDP \eqref{e:weight SDP dual final}. In particular, we see that unsteerable assemblages are those which violate no steering equalities in this form (they attain a value $ \mu^* = 1$, indicating a steerable weight $SW= 1-\mu^* = 0$), and that maximally steerable assemblages are those which maximally violate a steering inequality (obtaining value $ \mu^* = 0$, indicating a steerable weight $SW=1$). \dani{Finally, if one finds that a specific steering inequality (put in the form given in \eqref{e:weight SDP dual final}) is violated, this violation provides an upper bound on $\mu^*$, or equivalently a lower bound on the steerable weight $SW$.}


\subsection{D: All pure entangled states are maximally steerable}
In this section we will prove that all pure entangled states (of arbitrary dimension $d$) are steerable, and furthermore that it is sufficient to generate assemblages consisting of only $2d$ states (\ie by performing 2 $d$-outcome measurements). 

Our proof method relies on the dual characterisation of the steerable weight; from \eqref{e:weight SDP dual final} we know that maximally steerable assemblages are those which maximally violate a steering equality. Thus in order to prove that all entangled pure states are maximally steerable it suffices to show how one can always construct a steering inequality of the form needed for \eqref{e:weight SDP dual final} which is maximally violated.

Consider a pure entangled state of two qudits $\ket{\psi}$, written in its Schmidt decomposition as
\begin{equation}
\ket{\psi} = \sum_i \sqrt{p_i} \ket{i_a}\ket{i_b}
\end{equation}
Alice will perform two maximal projective von Neumann  measurements $\{M_{a|x}\}_{ax}$, with $x = 0,1$, and $a = 0,\ldots,d-1$, with each element a one-dimensional projector, $M_{a|x} = \ket{\phi_{a|x}}\bra{\phi_{a|x}}$. Alice creates the assemblage $\{\sigma_{a|x}\}_{ax}$, where $\sigma_{a|x} = p(a|x)\ket{\psi_{a|x}}\bra{\psi_{a|x}}$ with
\begin{equation}
\begin{aligned}
p(a|x) &= \sum_i p_i |\langle \phi_{a|x} | i_a\rangle |^2 \\
\ket{\psi_{a|x}} &= \frac{1}{\sqrt{p(a|x)}}\sum_i \sqrt{p_i} \langle \phi_{a|x} | i_a\rangle \ket{i_b}
\end{aligned}
\end{equation}
The key property of this assemblage is that each state contained in it is pure. This allows us to define the set of operators $\{F_{a|x}\}_{ax}$ for the steering inequality as 
\begin{equation}
F_{a|x} = \alpha(\openone - \ket{\psi_{a|x}}\bra{\psi_{a|x}})
\end{equation}
where $\alpha \geq 0$ is a constant to be determined later, each element has rank $d-1$, and, by construction, we have $\Tr F_{a|x}\sigma_{a|x} = 0$, from which it follows immediately that the assemblage attains the optimal value of 0 for this steering inequality. It remains to show however that the steering inequality, as constructed in this manner, always satisfies the conditions of \eqref{e:weight SDP dual final}. The final condition $F_{a|x} \geq 0$ is trivially satisfied, so in the remainder we will focus on the non-trivial first set of conditions.

The number of extremal single party deterministic strategies $D_\lambda(a|x)$ for the case of two $d$-outcomes measurements is $d^2$, with each distribution containing only two non-zero elements, corresponding to the choices of $a$ for $x = 0$ and $x=1$, each occurring with probability 1, which we shall refer to as $a_0$ and $a_1$ respectively. Thus, in the present context, the first set of conditions in \eqref{e:weight SDP dual final} reads
\begin{equation}\label{e:simplified conditions}
\openone - F_{a_0|0} - F_{a_1|1} \leq 0
\end{equation}
for all $d^2$ choices of $(a_0,a_1)$. Let us define $\xi$ as the maximum overlap between states of the assemblage when $x=0$ and $x=1$ are measured, that is
\begin{equation}
\xi = \max_{a_0,a_1} |\langle \psi_{a_0|0} | \psi_{a_1|1}\rangle|
\end{equation} 
when $\xi < 1$ then no state prepared for Bob by the first measurement is parallel to any state prepared for Bob by the second measurement. It follows therefore that the support of $F_{a_0|0} + F_{a_1|1}$ is the full space for all $a_0$ and $a_1$, and hence we can always choose an $\alpha$ sufficiently large such that the conditions \eqref{e:simplified conditions} are satisfied in all instances. 

Finally, $\xi < 1$ can be satisfied whenever the Schmidt rank of $\ket{\psi}$ is greater than 1 (\ie the state is entangled). As long as we ensure all $\ket{\phi_{a|x}}$ are non parallel on the support of $\{\ket{i_a}\}$ then the resulting states $\ket{\psi_{a|x}}$ will remain non parallel. 

In summary, the assemblages created by performing projective measurements on pure entangled states can always be shown to be maximally steerable by explicitly constructing a steering inequality from the assemblage. This construction should generalise beyond pure entangled states, to mixed states that prepare non-full rank states for Bob. Apart from the special case which we consider in the next section, we leave the exploration of this possibility for future work.  

\subsection{E: Local Hidden State model for $3 \times 3$ Werner state with MUBs}
In this section we give explicitly the LHS model which is able to reproduce the assemblage formed when Alice performs measurements of 4 mutually unbiased bases on the Werner state with $d = 3$ (see main text). We will denote the set of 4 MUBs by $\{M_{a|x}\}_{ax}$, with $x = 0,1,2,3$, $a = 0,1,2$, and each $M_{a|x} = |\phi_{a|x}\rangle\langle\phi_{a|x}|$. We will restrict our analisis to the specific set of MUBs given in 
\eqref{e:MUBs} with $\omega = e^{2\pi i/3}$, but a similar proof can be developed for any other set of MUBs (obtained by applying a rotation on  \eqref{e:MUBs}). We will again focus on the antisymetric projector\eqref{anti state} (\ie $\eta=1$), as a LHS model for the general Werner state (\ie $\eta\leq1$) can be  straighforwardly obtained from the one given here by simply mixing the identity, completely random, assemblage. 

From the previous subsection, we know that the assemblage formed by Alice when performing projective measurements on \eqref{anti state} will be given by 
\begin{equation}\label{e:sigma_ax}
\sigma_{a|x} = \tfrac{1}{2}(\openone_{3}-\ket{\phi_{a|x}}\bra{\phi_{a|x}}).
\end{equation}
where 
\begin{align}\label{e:MUBs}
\begin{aligned}
\ket{\phi_{0|0}} &= \ket{0}\\
\ket{\phi_{1|0}} &= \ket{1}\\
\ket{\phi_{2|0}} &= \ket{2}\\
\\ 
\ket{\phi_{0|1}} &= \tfrac{1}{\sqrt{3}}(\ket{0}+\ket{1}+\ket{2}) \\
\ket{\phi_{1|1}} &= \tfrac{1}{\sqrt{3}}(\ket{0}+\omega\ket{1}+\omega^2\ket{2}) \\
\ket{\phi_{2|1}} &= \tfrac{1}{\sqrt{3}}(\ket{0}+\omega^2\ket{1}+\omega\ket{2}) \\
\\
\ket{\phi_{0|2}} &= \tfrac{1}{\sqrt{3}}(\ket{0}+\omega\ket{1}+\omega\ket{2}) \\
\ket{\phi_{1|2}} &= \tfrac{1}{\sqrt{3}}(\ket{0}+\omega^2\ket{1}+\ket{2}) \\
\ket{\phi_{2|2}} &= \tfrac{1}{\sqrt{3}}(\ket{0}+\ket{1}+\omega^2\ket{2}) \\
\\
\ket{\phi_{0|3}} &= \tfrac{1}{\sqrt{3}}(\ket{0}+\omega^2\ket{1}+\omega^2\ket{2}) \\
\ket{\phi_{1|3}} &= \tfrac{1}{\sqrt{3}}(\ket{0}+\ket{1}+\omega\ket{2}) \\
\ket{\phi_{2|3}} &= \tfrac{1}{\sqrt{3}}(\ket{0}+\omega\ket{1}+\ket{2}) \\
\end{aligned}
\end{align}

This assemblage can be reproduced with the following LHS model, where the random variable $\lambda$ shared by Alice and Bob takes on one out of 9 values with equal probability $p(\lambda) = \frac{1}{9}$; conditioned on its value Alice has a deterministic value $a_x$ to output for each value of $x$, and Bob holds $\sigma_\lambda$, given by Table \ref{t:LHS model}.
\begin{table}
\begin{tabular}{|c|c|c|c|c|c|c|}
\hline 
$\lambda$ & $p(\lambda)$ & $a_0$ & $a_1$ & $a_2$ & $a_3$ & $\sigma_\lambda$ \\ 
\hline 
1 & $\tfrac{1}{12}$ & 0 & 0 & 0 & 0 & $\tfrac{1}{\sqrt{2}}\left(\ket{1}-\ket{2}\right)$ \\ 
\hline 
2 & $\tfrac{1}{12}$ & 0 & 1 & 1 & 1 & $\tfrac{1}{\sqrt{2}}\left(\omega^2\ket{1}-\ket{2}\right)$ \\ 
\hline 
3 & $\tfrac{1}{12}$ & 0 & 2 & 2 & 2 & $\tfrac{1}{\sqrt{2}}\left(\omega\ket{1}-\ket{2}\right)$ \\ 
\hline 
4 & $\tfrac{1}{12}$ & 1 & 0 & 1 & 2 & $\tfrac{1}{\sqrt{2}}\left(\ket{0}-\ket{2}\right)$ \\ 
\hline 
5 & $\tfrac{1}{12}$ & 1 & 1 & 2 & 0 & $\tfrac{1}{\sqrt{2}}\left(\omega\ket{0}-\ket{2}\right)$ \\ 
\hline 
6 & $\tfrac{1}{12}$ & 1 & 2 & 0 & 1 & $\tfrac{1}{\sqrt{2}}\left(\omega^2\ket{0}-\ket{2}\right)$ \\ 
\hline 
7 & $\tfrac{1}{12}$ & 2 & 0 & 2 & 1 & $\tfrac{1}{\sqrt{2}}\left(\ket{0}-\ket{1}\right)$ \\ 
\hline 
8 & $\tfrac{1}{12}$ & 2 & 1 & 0 & 2 & $\tfrac{1}{\sqrt{2}}\left(\omega^2\ket{0}-\ket{1}\right)$ \\ 
\hline 
9 & $\tfrac{1}{12}$ & 2 & 2 & 1 & 0 & $\tfrac{1}{\sqrt{2}}\left(\omega\ket{0}-\ket{1}\right)$ \\ 
\hline 
\end{tabular} 
\caption{Local Hidden State model for MUB measurements on $3 \times 3$ Werner state. $\lambda$ labels the hidden variable; $p(\lambda)$ gives the probability for a given $\lambda$; $a_{x}$ gives the deterministic output Alice will give for each $x$, given the variable $\lambda$; and $\sigma_\lambda$ is the state prepared for Bob for each $\lambda$.}\label{t:LHS model}
\end{table}

The important point to note is that for a given input and outcome, $\tilde{x}$ and $\tilde{a}$ respectively, there are always precisely 3 values of $\lambda$ which are compatible (\ie there are 3 values of $\lambda$ such that this outcome will occur with non-zero probability). The 3 states $\sigma_\lambda$ which correspond to these values of $\lambda$ are all orthogonal to the state $\ket{\phi_{a|x}}$ and when mixed with equal probability form exactly the state \eqref{e:sigma_ax} which is the projector onto the orthogonal subspace of $\ket{\phi_{a|x}}$.


\subsection{F: The antisymmetric state is maximally steerable}

In this subsection we show that  the state defined as the normalized antisymetric projector, \ie the Werner state with $\eta=1$, has steerable weight equal to one. The demonstration will basically follow the same one for entangled pure states given in the previous section: we will show that this state maximally violates a steering inequality.

First notice that the Werner state for $\eta=1$ can be written as
\begin{eqnarray}\label{anti state}
\rho^d_W(\eta=1)&=&\frac{2A_d}{d-1}=\frac{\openone_{d^2}-F}{d-1},
\end{eqnarray}
where $F=\sum_{i,j=0}^{d-1}\ket{ij}\bra{ji}$ is the flip operator. Consider again that Alice applies projective von Neumann  measurements $\{M_{a|x}\}_{ax}$ ($a = 0,\ldots,d-1$), each measurement labelled by $x$, on the state \eqref{anti state}, with each element a one-dimensional projector, $M_{a|x} = \ket{\phi_{a|x}}\bra{\phi_{a|x}}$. The assemblage generated in Bob's site will be the given by
\begin{eqnarray}\label{anti assemblage}
\sigma_{a|x}&=&\Tr_A(\ket{\phi_{a|x}}\bra{\phi_{a|x}}\otimes\openone_{d} \times\frac{\openone_{d^2}-F}{d-1})\nonumber\\
&=&\frac{1}{d-1}[\openone_{d}-\Tr_A(\ket{\phi_{a|x}}\bra{\phi_{a|x}}\otimes\openone_{d}\times \sum_{i,j=0}^{d-1}\ket{ij}\bra{ji})]\nonumber\\
&=&\frac{1}{d-1}(\openone_{d}-\sum_{i,j=0}^{d-1}\braket{j}{\phi_{a|x}}\braket{\phi_{a|x}}{i}\ket{j}\bra{i})\nonumber\\
&=&\frac{1}{d-1}(\openone_{d}-\ket{\phi_{a|x}}\bra{\phi_{a|x}}).
\end{eqnarray}
As before, we can define a set of operators $\{F_{a|x}\}_{ax}$ for the steering inequality:
\begin{equation}
F_{a|x} = \alpha\ket{\psi_{a|x}}\bra{\psi_{a|x}}
\end{equation}
where $\alpha \geq 0$ is a constant to be determined later, each element has rank $1$, and, we have $\Tr F_{a|x}\sigma_{a|x} = 0$. It then follows that the assemblage \eqref{anti assemblage} attains the optimal value of 0 for this steering inequality.  It remains now to show that there will be a choice of $\alpha$ and vectors $\ket{\phi_{a|x}}\bra{\phi_{a|x}}$ which can be used to construct of a proper steering inequality, \ie such that $\{F_{a|x}\}_{a,x}$ satisfy the constraints imposed in \eqref{e:weight SDP dual final}. The condition $F_{a|x} \geq 0$ is trivially satisfied, so in the remainder we will focus on the non-trivial first set of conditions. 

First notice that what the first set of constraints in \eqref{e:weight SDP dual final} is saying is that if one makes the summation of one element from each basis this should result in an operator bigger than the identity operator ($\lambda$ labels which combination of elements). We will guarantee this by choosing bases $\{\ket{\phi_{a|x}}\}_{a,x}$ with the property that the set of vectors composed by one element from each of these bases always span the whole space. Then, by choosing $\alpha$ big enough we can guarantee that the conditions \eqref{e:simplified conditions} are satisfied in all instances.

We will argue that $d$ random bases fulfil this property. First choose, for $x=0$, a random a basis $\{\ket{\phi_{a|0}}\}_{a=0,..,d-1}$. Then do the same for $x=1$. The probability that one of the elements of the new basis $\{\ket{\phi_{a|1}}\}_{a=0,..,d-1}$ is parallel to any one of the elements of the previous basis $\{\ket{\phi_{a|0}}\}_{a=0,..,d-1}$ is zero. We can now generate a new random basis, for $x=2$. The probability that this new basis contains a vector that lies in a plane generated by any pairs of vector from the bases corresponding to $x=0$ and $x=1$, \ie $\ket{\phi_{a|2}}=a\ket{\phi_{i|0}}+b\ket{\phi_{j|1}}$, is zero. By repeating this argument until $x=d-1$ we can see that the support of 
\begin{equation}
F_{a_0|0}+F_{a_1|1}+\ldots+F_{a_{d-1}|d-1}
\end{equation}
is the full space for any choices of $a_0,\ldots,a_{d-1}$. Then, by choosing $\alpha$ big enough we can guarantee that
\begin{equation}
\openone \leq \sum_{ax} D_\lambda(a|x)F_{a|x} \quad \forall \lambda
\end{equation}
as required in \eqref{e:weight SDP dual final}.

\subsection{G: One way steering of the erasure state}
In this section we show that, for any number of general POVM measurements $k$ in Alice's side, there exists an erasure state that is unsteerable from Alice to Bob but the same state is steerable from Bob to Alice, with Bob needing only to perform two projective measurements. 

The qubit erasure state $\rho_p^\mathrm{er}$ is given by
\begin{equation}\label{e:erasure state}
\rho_p^\mathrm{er} = p \ket{\psi^-}\bra{\psi^-} + (1-p)\ket{2}\bra{2}\otimes \openone/2
\end{equation}
i.e. it is a qutrit-qubit state 
and is so called as it is the state which results from sending Alice's half of the singlet $\ket{\psi^-}$ through the erasure channel with parameter $p$ and flag state $\ket{2}$.

Our first goal is to show that there is a LHS model for any assemblage which arises from $k$ POVM measurements (with an arbitrary number of outcomes) performed by Alice on the erasure state, when $p \leq 1/k$. This follows from the fact that for $p \leq 1/k$ the state has a $k$--symmetric extenstion on Alice's side, meaning that there exists an extension $\rho^\mathrm{ext}_{A_1\cdots A_k B} \in \mathbb{C}^3 \otimes \cdots \otimes \mathbb{C}^3 \otimes \mathbb{C}^2$ such that (i) $\rho^\mathrm{ext}_{A_iB} \equiv \Tr_{\bar{A_i}} \rho^\mathrm{ext}_{A_1\cdots A_k B} = \rho_p^\mathrm{er} $, where $\bar{A_i}$ denotes all parties $A_j \neq A_i$; (ii) $F_{A_i A_j} \rho^\mathrm{ext}_{A_1\cdots A_k B} F_{A_i A_j} = \rho^\mathrm{ext}_{A_1\cdots A_k B}$ for all $i \neq j$, where $F_{A_i A_j}$ is the operator which acts as \textsc{swap} between $A_i$ and $A_j$ and as the identity on the rest of the space. The $k$-symmetric extension of 
\eqref{e:erasure state} is
\begin{equation}
	\rho^\mathrm{ext}_{A_1\cdots A_k B} = \frac{1}{k} \sum_{i} \ket{2}\bra{2}^{\otimes (k-1)}_{\bar{A}_i} \otimes \ket{\psi^-}\bra{\psi^-}_{A_i B}
\end{equation}
which is easily seen to satisfy properties (i) and (ii). Intuitively the existence of a $k$--symmetric extension implies a LHS model for $k$ measurements on Alice since Alice can perform each of her $k$ measurements on a different copy of the extension, thus meaning she actually needs to only perform a single measurement, which always has a LHS model. More precisely, for the collection of $k$ POVM measurements $M_{a|x}$, for $x = 0,\ldots,k-1$, and $a = 0,\ldots, m-1$, for arbitrarily $m$, define the single POVM 
\begin{equation}
M_{\mathbf{a}} = Ṃ_{a_0|0} \otimes \cdots \otimes Ṃ_{a_{k-1}|k-1}
\end{equation}
where $\mathbf{a} = a_0 a_1\cdots a_{k-1}$ is a string containing the outputs of all $k$ measurements. Now define $p(\mathbf{a})$ as
\begin{equation}
p(\mathbf{a}) = \Tr \left(M_{\mathbf{a}} \rho^\mathrm{ext}_{A_1\cdots A_k}\right)
\end{equation}
where $\rho^\mathrm{ext}_{A_1\cdots A_k} = \Tr_B \rho^\mathrm{ext}_{A_1\cdots A_k B}$ is the reduced state of the extension on the Alices. This probability distribution has the property that
\begin{equation}\label{e:correct p}
\sum_{a_j \neq a_i} p(\mathbf{a}) = p(a|x = i) = \Tr \left(\left(M_{a|i} \otimes \openone_B\right)\rho_p^\mathrm{er}\right)
\end{equation}
Similarly, define the sub-normalised states $\sigma_\mathbf{a}$ as
\begin{equation}
 \sigma_\mathbf{a} = \Tr_{A_1\cdots A_k} \left(\left(M_{\mathbf{a}} \otimes \openone_B\right) \rho^\mathrm{ext}_{A_1\cdots A_k B}\right)
\end{equation}
which have the analogous property that 
\begin{equation}\label{e:correct sig}
\sum_{a_j \neq a_i} \sigma_\mathbf{a} = \sigma_{a|x=i} = \Tr_A \left(\left(M_{a|i}\otimes \openone_B\right) \rho_p^\mathrm{er}\right)
\end{equation}
The local hidden state model amounts then to sending $\lambda = \mathbf{a}$ to Alice and $\sigma_\lambda = \frac{1}{p(\mathbf{a})}\sigma_\mathbf{a}$ to Bob with probability $p(\lambda) = p(\mathbf{a})$, with Alice's response function being that upon receiving $x$ she outputs $a_x$ from her string. Since there are exactly $m^k$ different strings $\mathbf{a}$, we see that each $\mathbf{a}$ in fact labels a deterministic strategy $D_\mathbf{a}(a|x)$ for Alice. Finally, to see that this LHS model faithfully reproduces the assemblage $\sigma_{a|x} = \Tr_A \left(\left(M_{a|x}\otimes \openone_B\right)  \rho_p^\mathrm{er} \right)$, we see from \eqref{e:correct p} that Alice outputs $a$ with the correct probability for all $x$, and similar from \eqref{e:correct sig} that Bob holds the correct state from the assemblage in each case. 

In the converse direction, we now want to show that if instead it is Bob who wishes to steer to Alice, 
then for all $p>0$ the state is steerable. To do we show that we can generate an assemblage that always violates a steering inequality, found by solving the dual SDP (12). 

To that end, consider that Bob performs the two Pauli measurements $X$ and $Z$ on the erasure state, which generates the assemblage $\sigma_{b|y}$
\begin{align}\label{e:assemblage erasure}
\begin{aligned}
	\sigma_{0|0} &= \tfrac{p}{2}\ket{\downarrow_x}\bra{\downarrow_x} + \tfrac{(1-p)}{2}\ket{2}\bra{2} \\
	\sigma_{1|0} &= \tfrac{p}{2}\ket{\uparrow_x}\bra{\uparrow_x} + \tfrac{(1-p)}{2}\ket{2}\bra{2} \\
	\sigma_{0|1} &= \tfrac{p}{2}\ket{\downarrow_z}\bra{\downarrow_z} + \tfrac{(1-p)}{2}\ket{2}\bra{2} \\
	\sigma_{1|1} &= \tfrac{p}{2}\ket{\uparrow_z}\bra{\uparrow_z} + \tfrac{(1-p)}{2}\ket{2}\bra{2} \\
\end{aligned}
\end{align}
Consider also the steering inequality with elements $F_{b|y}$
\begin{align}\label{e:erasure ineq}
\begin{aligned}
	F_{0|0} &= \alpha\ket{\uparrow_x}\bra{\uparrow_x} + \tfrac{1}{2}\ket{2}\bra{2} \\
	F_{1|0} &= \alpha\ket{\downarrow_x}\bra{\downarrow_x} + \tfrac{1}{2}\ket{2}\bra{2} \\
	F_{0|1} &= \alpha\ket{\uparrow_z}\bra{\uparrow_z} + \tfrac{1}{2}\ket{2}\bra{2} \\
	F_{1|1} &= \alpha\ket{\downarrow_z}\bra{\downarrow_z} + \tfrac{1}{2}\ket{2}\bra{2} \\
\end{aligned}
\end{align}
where $\alpha = 2 + \sqrt{2}$. It is easy to check that $F_{b|0}~+~F_{b'|1}~\geq~\openone$ for all $b$,$b'$, hence from (12) they constitute a valid steering inequality for which all unsteerable assemblages take value at least 1. However, the assemblage \eqref{e:assemblage erasure} takes the value
\begin{equation}
\Tr\sum_{b,y} F_{b|y} \sigma_{b|y} = (1-p)
\end{equation}
Hence for all $p>0$ the assemblage violates the steering inequality and thus demonstrates steering. Intuitively the fact that the erasure state is steerable in this direction follows simply from the fact that a trusted Alice can determnie if she holds the flag state, and, upon not seeing it, she knows for sure that they hold a singlet state. 

Put together, the erasure state with $p = 1/k$ therefore provides an example of a state which on the one hand (when the untrusted party holds the flag), for $k$ or less measurements there is a local hidden state model for all POVM measurements, with arbitrary numbers of outcomes. On the other hand, when the trusted party holds the flag, using only two projective measurements, steering can be demonstrated by violating the inequality \eqref{e:erasure ineq}. Only in the limit $k \to \infty$ does the construction fail to work, as it is known that states with an infinite symmetric extension is separable.

\end{appendix}

\end{document}